\documentclass{llncs}

\usepackage{rotating}
\usepackage{booktabs,fancyhdr} 

\usepackage[round,authoryear]{natbib}

\usepackage{amsmath}
\usepackage{amssymb}
\usepackage{graphicx}

\newcounter{xxx}
\def\bbP{\mathbb{P}}
\newcommand\identitystates{{\mathcal S}}

\title{Correcting for Cryptic Relatedness in Genome-Wide Association Studies}
\titlerunning{Correcting for Cryptic Relatedness in GWAS}

\author{Bonnie Kirkpatrick\inst{1,*} and
Alexandre Bouchard-C\^ot\'e\inst{2}
}
\institute{Intrepid Net Computing, \email{bbkirk@intrepidnetcomputing.com}, 406-998-0179, Montana, US, *corresponding author,
\and University of British Columbia, Statistics, \email{bouchard@stat.ubc.ca}, 
605-822-1669, 3182 Earth Sciences Building, 2207 Main Mall, Vancouver, BC, Canada V6T 1Z4.}

\begin{document}
\maketitle

\begin{abstract}
While the individuals chosen for a genome-wide association study (GWAS) may not be closely related to each other, there can be distant (cryptic) relationships that confound the evidence of disease association.  These cryptic relationships violate the GWAS assumption regarding the independence of the subjects' genomes, and failure to account for these relationships results in both false positives and false negatives.

This paper presents a method to correct for these cryptic relationships.  We accurately detect distant relationships using an expectation maximization (EM) algorithm for finding the identity coefficients from genotype data with know prior knowledge of relationships.  From the identity coefficients we compute the kinship coefficients and employ a kinship-corrected association test.

We assess the accuracy of our EM kinship estimation algorithm. We show that on genomes simulated from a Wright-Fisher pedigree, our method converges quickly and requires only a relatively small number of sites to be accurate. We also demonstrate that our kinship coefficient estimates outperform state-of-the-art covariance-based approaches and PLINK's kinship estimate.

To assess the kinship-corrected association test, we simulated individuals from  deep pedigrees and drew one site to recessively determine the disease status.  We applied our EM algorithm to estimate the kinship coefficients and ran a kinship-adjusted association test.  Our approach compares favorably with the state-of-the-art and far out-performs a na\"ive association test.

We advocate use of our method to detect cryptic relationships and for correcting association tests.  Not only is our model easy to interpret due to the use of identity states as latent variables, but also inference provides state-of-the-art accuracy.

\end{abstract}

\section{Introduction}

%The classical way to correct for relationship structure in a disease association study relies on the full knowledge of the relationships in the form of a pedigree graph, and to consider that graph when computing the association.  These methodologies are referred to as linkage analysis, and known algorithms in this category are exponential in the number of individuals or sites,~\cite{Elston1971,Lander1987,Abecasis2002,Lauritzen2003,Piccolboni2003}.  
%Computing the pedigree likelihood has been shown to be NP hard~\cite{Piccolboni2003} and the likelihood plays a critical role in exact linkage analysis.  
%For the reason that the pedigree may be very large~\cite{Abney2002,Sutter2007} and that relationships may be unknown, it is often not practical to use linkage analysis in disease association studies.

Accuracy in disease association studies is heavily influenced by cryptic relatedness and population substructure,~\cite{Astle2009}.  Both false positives and false negatives result from these influences, because of false assumptions of independence between individuals that are actually related.  Disease associations are primarily detected using the genome-wide association study (GWAS) which is typically a case-control or cohort association study implemented using a test for correlation between the disease and the genotypes,~\cite{Risch1996}.  While GWAS typically assumes independence between the individuals, a growing number of methods are designed to detect relationships in the form of statistical dependence between the genomes of individuals and correct the correlation calculation for these dependencies.  This paper presents a novel method for detecting relationships and correcting the association analysis.

Methods exist that correct association test for relationships that are either known or unknown.  These methods often use a two-step approach.  First, pair-wise relationships in the form of identity states or kinship coefficients must be inferred either from a known pedigree structure,~\cite{thompson1985}, or from data,~\cite{Purcell2007,Browning2011,Milligan2003,Sun2014}.  Second, the inferred relationships are used to correct the test for association,~\cite{Yu2005,Thornton2007,Rakovski2009,Kang2010,Cordell2014}.

In this work, we focus on inferring pair-wise relationships, for which existing methods can be split into four categories.  First, there are exact methods that compute the kinship coefficients from pedigree structures in quadratic time in the number of individuals in the pedigree,~\cite{Kirkpatrick2016a,Abney2009,thompson1985,Karigl81}.  Second, there are ways to estimate the probabilities of the condensed identity coefficients,~\cite{Jacquard1972}, from data using a maximum likelihood estimator,~\cite{Milligan2003}.  
%The kinship coefficients are simply an expectation with respect to the condensed identity coefficients, and can be readily computed from these estimates.  
Third, there are ways to estimate the probabilities of the outbred condensed identity coefficients,~\cite{Purcell2007,Browning2011}\footnote{The probabilities of having zero, one, or two alleles identical-by-descent (IBD) are considered.}, but such methods work only for outbred pedigrees and fail to capture more complicated relationships, as we see later in this paper.
Forth, there has been work on inferring kinship coefficients from admixed individuals,~\cite{Thornton2012}.
%Fourth, the is a method called REAP for inferring kinship coefficients from admixed individuals~\cite{Thornton2012}.  However, in our tests REAP was unable to make reasonable inferences about independent individuals from the same population.  %% moving this to the experiments section

Our EM method has a running time, for a pair of individuals, that is linear in the number of sites.  To estimate a kinship coefficient matrix for $n$ individuals, our method is quadratic in $n$, due to the number of pairs of individual. Our method avoids the assumption of having a known pedigree, taking only the genotypes as an input. We introduce one latent variable for each pair of individuals and each site, which encodes our uncertainty on the identity by descent states. The distribution over these states for each pair of individuals is learned using a simple expectation maximization (EM) algorithm and provides an informative, yet concise and learnable, summary of the relationship of each pair of individuals. 
%This updated distribution is then used to re-estimate the posterior distribution over the latent variables. This process is repeated until convergence.

We show that this EM algorithm quickly converges to an accurate estimate. We assess the accuracy in two ways.  First, we measure the accuracy of the reconstructed latent variables using simulated pedigree data (the true value of the latent variable can be computed efficiently from the held-out pedigree structure). We asses the kinship estimates compared with PLINK and with a covariance-based estimate.  Second, we demonstrate that our method can be used to correct GWAS for cryptic relatedness without assuming the knowledge of the pedigree structure.

\section{Background}\label{sec:background}

\begin{figure}[ht!]
  \begin{center}
  \includegraphics[width=5in]{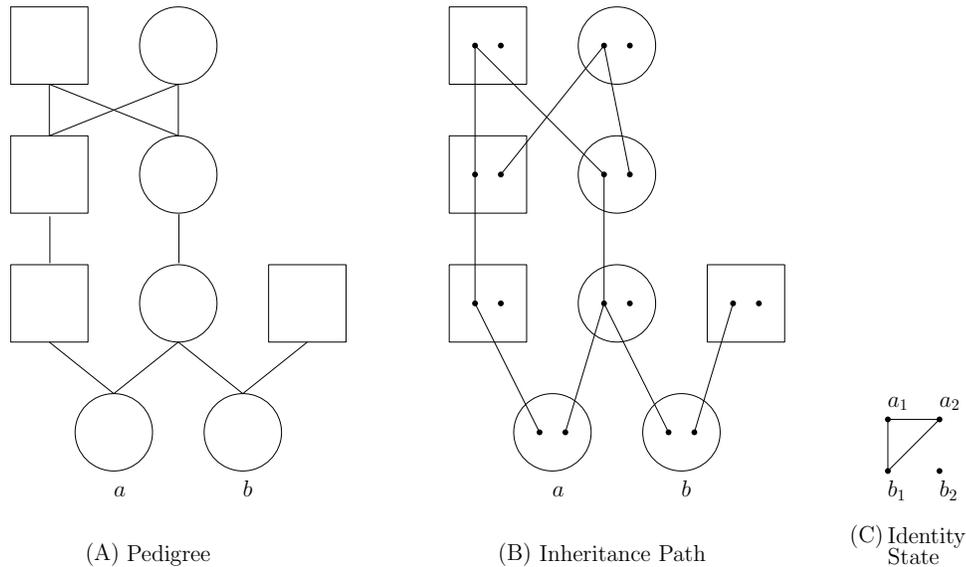}
  \end{center}
  \caption{{\bf Example pedigree, inheritance path, and identity state.} {\bf A)} An example of a pedigree dag is shown at the left with the edges directed implicitly downwards, the circles representing females, and the boxes representing males.  The two individuals with known genotypes are labeled $a$ and $b$.  {\bf B)} An example of an inheritance path graph for this pedigree is shown in the center with the dots being nodes representing alleles.  Although the allele nodes are unlabeled, here, for each individual the inheritance path edges are confined to the options permitted by the pedigree graph.  Notice that while there are cycles in graph (A), the inheritance path graph, which has alleles (dots) as nodes, has no cycles (such a graph is called a forest, as it is not connected in general).  There are three trees in this inheritance path forest.  
 %\XXX{Since there are cycle, shouldn't we call these DAGs instead of trees? If so, the same change should be applied to the Background section and perhaps other places too. It depends on what are the nodes of the graph, and the inheritance path is acyclic.  Is this clear enough, or do we need to add more explanation?}  
{\bf C)} An example of an identity state is shown on the right.  This identity state is the one produced from the connected components of the inheritance path graph in the center (B).}
  \label{fig:ped-ibd}
\end{figure}

A pedigree is an annotated directed acyclic graph where the nodes are diploid individuals and the edges are directed from parent to child (Figure~\ref{fig:ped-ibd}A).  The graph is acyclic, because no individual can be their own ancestor.  
The pedigree graph is typically annotated with the gender (male or female), affection status, and the genotype status of the individuals.  The gender is used to enforce on the graph the marriage constraint that each individual with parents in the graph has at most one parent of each gender. The \emph{founders} are the individuals who do not have parents in the pedigree while \emph{non-founders} are all the individuals having parent(s) in the pedigree graph.  The \emph{affection status} indicates whether an individual has a disease or phenotype.  The genotype status indicates whether the individual has available genotypes.

The pedigree graph depicts familial relationships between individuals.  Each diploid individual inherits one copy of each chromosome from each parent, and at different sites (loci) in the genome, an individual may inherit from different copies of the parent's chromosome.  In representing inheritance, we consider a single site in the genome and the binary inheritance decision made along each parent-child edge in the pedigree graph. We represent this with an \emph{inheritance path} (Figure~\ref{fig:ped-ibd}B).  For a particular genomic site, the nodes of the inheritance path represent the alleles in all the individuals in the pedigree graph (i.e.~for individual $a$ the allele nodes are $a_1$ and $a_2$).  
The edges of the inheritance path graph proceed from a parent allele to a child allele and indicate that the parent allele is copied with fidelity (i.e.~without mutation) to the child allele.  
%This graph is a forest, because each non-founder allele has exactly one parental source.  
For each pedigree graph with $n$ edges there are $2^n$ possible inheritance path graphs. 

In an inheritance path, if two alleles are in the same connected component, we say those alleles are \emph{identical by descent (IBD)}.  For example, the paternal alleles from two siblings are IBD if and only if the two children inherit from the same allele of the father's two alleles.  The two alleles from the same individual whose parents are related to each other may appear in the same connected component due to inbreeding.

When we focus on two individuals, $a$ and $b$, who are genotyped, a natural question is: which subsets of the four alleles of those two individuals  are IBD? For a given inheritance path, the \emph{identity state} captures the answer to this question (Figure~\ref{fig:ped-ibd}C).  The identity state is a graph with four nodes, one for each of the four alleles $\{a_1,a_2,b_1,b_2\}$, and with an edge between two alleles if and only if these two alleles are IBD.  Therefore, there is an edge between two alleles in the identity state if and only if they appear in the same connected component of the inheritance path graph.

The identity state is related to the genotype data in the following way.  Each of these two individuals may be either heterozygous (two different alleles, i.e. $a_1 \ne a_2$) or homozygous (two copies of the same allele, $a_1 = a_2$) at each site assayed.  Certainly if alleles $x,y$ appear together in a connected component of the identity state, they are IBD and must be identical alleles.  However, if alleles $x,y$ appear in different connected components of the identity state, then they are not IBD and they may or may not be identical alleles.

For any given pedigree, there are 15 possible identity states,~\cite{Jacquard1972}.  For a pedigree with $n$ edges and a pair of individuals $a$ and $b$, the identity state is a random variable, $S$, having a corresponding distribution called the \emph{identity coefficients}, $\bbP[S=s]$, which is the fraction of the $2^n$ inheritance paths for the pedigree that exhibit identity state $s$.

\begin{figure}[ht!]
  \begin{center}
  \includegraphics[width=3in]{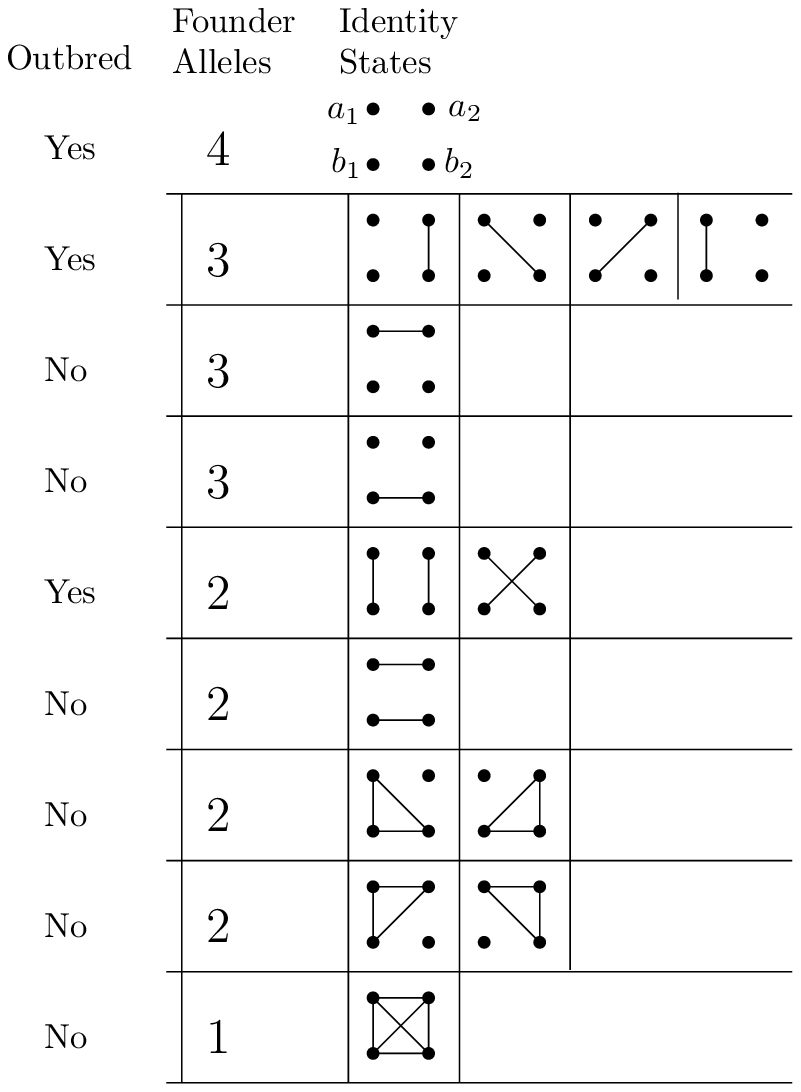}
  \end{center}
  \caption{{\bf Identity States.} The 15 identity states are grouped so that each row corresponds to one of the 9 condensed identity states.  The number of founder alleles for each identity state is listed, along with whether the identity state is outbred.
}
  \label{fig:identstates}
\end{figure}

While the identity state describes the possible IBD between two genotyped individuals, the \emph{condensed identity state},~\cite{Jacquard1972}, more directly relates the IBD to the genotype data.  In the identity state the two alleles of one individual are distinguishable, however in the genotype, due to our inability to measure which allele is on which chromosome, the alleles are exchangeable.  The condensed identity states are a grouping or partition of the identity states such that two identity states appearing in the same group are isomorphic when the labels on the nodes belonging to individual $a$ are permuted and the labels on the nodes belonging to the individual $b$ are permuted (any valid permutation is permitted, including the permutation that does not change the node labels of the individual).  According to this definition, there are nine condensed identity states.  Let $c_i$ be the number of condensed identity states which contain $i$ distinct identity states.  Then the 9 condensed identity states group the 15 identity states into groups $c_1=5, c_2=3, c_3=0, c_4=1$ (see Figure~\ref{fig:identstates} for drawings of all the identity state graphs grouped into the condensed identity states).  Similar to the identity coefficient, each of these condensed identity states has a corresponding \emph{condensed identity coefficient} which is the fraction of the $2^n$ inheritance paths for the pedigree with $n$ edges that exhibit the condensed identity state.

We need one last concept, still defined with respect to a pair of individuals in a pedigree.  The \emph{kinship coefficient} for that pair of individuals is defined as the probability of IBD when choosing one allele from each individual uniformly at random.  The exact kinship coefficients for every pair of individuals are typically computed from known relationships--a pedigree graph--using a recursive computation,~\cite{Kirkpatrick2016a,Abney2009,Karigl81,thompson1985}.

\section{Methods}\label{sec:methods}
\label{em}

We focus on estimating the identity coefficients.~\footnote{One could re-express our method in terms of the condensed identity coefficients (since the data do not provide information to distinguish between identity states that fall in the same condensed identity states), but the presentation is simpler in the non-condensed setting.}  
We do not directly estimate the kinship coefficients from the genotypes, because we have no model by which the kinship coefficients can generate the genotypes.  On the other hand, if we know the identity states, their coefficients, and the allele frequencies in the founders, there is a generative model for producing the genotype.  Since the kinship coefficients can be expressed as an expectation with respect to the identity coefficients, they contain a degenerate version of the information contained in the identity coefficients and do not provide a convenient generative model for the genotype.

The estimated identity coefficients are then transformed into estimates of the kinship using equations developed by~\cite{Kirkpatrick2011xxxx,Kirkpatrick2016a}, which we repeat for convenience.  For an identity state, let $t \in \{aa, ab, bb\}$ denote an \emph{edge type}, for example, $ab$ indicates any edge between the alleles in two different individuals $a$ and $b$ (i.e., and edge between one of the nodes $\{a_1, a_2\}$ and one of the nodes $\{b_1,b_2\}$). Let $\identitystates$ be the set of all the identity states for a pedigree and pair of individuals.  Let $e(s,t)$ for $s \in \identitystates$ be an indicator function that is one if and only if the identity state has an edge of type $t$.  Equations~(\ref{eq:phi_ab}) and (\ref{eq:phi_aa}) give the kinship in terms of the identity state distribution.
\begin{equation} \label{eq:phi_ab}
\Phi_{a,b} = \sum_{s\in\identitystates} \frac{e(s,ab)}{4} \; \bbP[S=s].
\end{equation}
where $s$ is the identity state.  Rather than the kinship coefficient for the diagonal, we consider the inbreeding coefficient~(see~\cite{Kirkpatrick2016a} for details of how these quantities are related):
\begin{equation} \label{eq:phi_aa}
\Phi_{a,a} = \sum_{s\in\identitystates} e(s,aa) \; \bbP[S=s],
\end{equation}
where $e(s,aa)$ indicates whether the single possible edge between nodes $a_1$ and $a_2$ exists.  

In this section, we first introduce a model with the identity states as latent variables.  Second, we discuss our EM algorithm for doing inference with this model.  Accuracy discussions appear in the Results section.

\subsection{Model}

% If we know the identity states and their coefficients and the allele frequencies in the founders, there is a generative model for producing the genotype.  Those estimated identity coefficients can later be transformed into estimates of the kinship.  We focus here on the identity states as latent variables, because we do not know of a model for generating genotypes from the kinship coefficients without also involving the identity coefficients.

Let the identity states for the $m$ sites be represented as $S = (S_1, \dots,S_m)$.  Let the pair genotype data be represented as $G = (G_1, \dots,G_m)$ where each $G_j$ is a tuple of two genotypes $G_j=(G^1_j,G^2_j)$, one genotype for each individual of interest.  For each genotype, the values $G^i_j$ takes values in $\{0,1,2\}$, the count of the minor allele.  The vector $p = (p_1, \dots, p_m)$ contains the founder allele frequencies for each site.

In our model, the likelihood is a function of the allele frequencies and is defined as a marginal probability where the identity states are marginalized out
\[
  \bbP[G=g|p] = \sum_{s\in\identitystates} \bbP[G=g,S=s|p].
\]
The joint probability of the data and the identity states is defined as the product of independent sites:\footnote{In practice, sites are clearly dependent because of recombination, but we assume sites have been sufficiently subsampled to approximate independence. We show in the next section that our method requires relatively few sites, so this is a reasonable approximation in practice.}
\begin{eqnarray}
  \bbP[G=g,S=s|p] &=& \prod_j \bbP[G_j=g_j,S_j=s_j|p_j] \\ 
              &=& \prod_j \bbP[G_j=g_j|S_j=s_j,p_j]\; \bbP[S_j=s_j].
\end{eqnarray}
Since all of the inheritance paths are drawn from the same pedigree, the distribution on the identity states is the same for all sites: $\bbP[S_j=s] = \bbP[S_{j'}=s]$ for all $s$.   We denote the parameters of this shared categorical distribution by $(d_s : s\in\identitystates)$, i.e. $d_s = \bbP[S_j=s]$.

A key component of our model is the conditional probability $\bbP[G_j=g|S_j=s,p_j]$.  To write an expression for this conditional probability, we introduce the \emph{allele assignment} which must be consistent with both the genotype and the identity state.  Recall that an identity state $s$ has vertices $V = \{a_1,a_2,b_1,b_2\}$.  Let an allele assignment be given by a map $a:V \to \{0,1\}$.  We say that an allele assignment $a$ with $1$ as the minor allele is consistent with the genotypes for two individuals $g=(g^1,g^2)$ if and only if $a(a_1)+a(a_2) = g^1$ and $a(b_1)+a(b_2) = g^2$.  Indicate this genotype consistency by the indicator function $C$ (i.e., $C(a,g) = 1$, otherwise $C(a,g) = 0$).  Further let $CC(s)$ be the partition of $V$ into connected components extracted from the identity state graph of $s$.  Then we say that an allele assignment $a$ is legal with respect to the identity state if and only if the function $a$ is constant on each connected component (i.e., for each connected component $c \in CC(s)$, $a(x) = a(y)$ whenever $x,y \in c$).  We represent a legal allele assignment with the indicator $L(a,s)=1$ and $L(a,s)=0$ otherwise.  Now, let 
\[A(s,g) = \{a | C(a,g)=1 \textrm{ and } L(a,s)=1\}\]
be the set of legal and consistent assignments of the genotype alleles to the allele nodes of the identity state.

We can now write an expression for the conditional probability $\bbP[G_j=g|S_j=s,p_j]$ which is:
\[
  \bbP[G_j=g|S_j=s,p_j] \propto \sum_{a \in A(s,g)} p_j^{n_0(a)} (1-p_j)^{N_{cc}-n_0(a)},
\]
where $A(s,g)$ is the set of legal assignments of the genotype alleles to the allele nodes of the identity state, $N_{cc}$ is the number of connected components in the identity state, $n_0(a)$ is the number of connected components of the identity state that are labeled with the minor allele by assignment $a$.

\subsection{Inference}

For inference, we use the EM algorithm below.  This algorithm takes as input an estimate of the allele frequencies $\hat{p}$ together with the genotypes for a pair of individuals and predicts the identity state coefficients for those individuals.  First, we will describe a method to estimate $\hat{p}$ from many genotypes, and second, we will give the details of the EM estimator.

For simplicity and computational efficiency, we estimate the allele frequencies $\hat{p}$ using the Laplace estimator based on the genotypes of independent individuals.\footnote{Smoothing of $\hat{p}$ is required to avoid degeneracies where our method fails due to divisions by zero in Equation~(\ref{eq:ml-eq}).}  
Since our simulations produce pairs of related individuals, we consider the genotypes of one individual per simulated pair, but pool the data from many pairs.
Later, in our simulations, this same allele frequency estimate $\hat{p}$ is then given to all the methods that infer kinship, ours and others.% the kinship coefficient estimation methods and to the EM algorithm described in the Methods section. 

To estimate the identity state coefficient, we use an EM algorithm that iteratively produces successive estimates $d_s^{(t)}$ of the probability distribution $\bbP[S_j=s]$ from above.  These estimates are obtained using simple and efficient update rules.  For the {\bf E-step} we update our estimate of $N^{(t)}_s$ which is the expected number of times that identity state $s$ occurred 
\begin{align}\label{eq:ml-eq}
N^{(t)}_s = \sum_{j} \frac{d_s^{(t-1)} \times \bbP[G_j=g_j|S_j=s,\hat{p}]}{\sum_{s'} d_{s'}^{(t-1)} \times \bbP[G_j=g_j|S_j=s',\hat{p}]}
\end{align}
while the {\bf M-step} consists of updating our estimate of the identity coefficients 
\[
d_s^{(t)} = \frac{N^{(t)}_s}{m}.
\]
Iterative application of the E-step and M-step yields a sequence of estimates:
\[(N^{(t)}_s,d_s^{(t)})~~\textrm{for } t=1,2,\dots~.\]
%To iterate the E-step uses $d_s^{(t-1)}$ to estimate $N^{(t)}_s$, and the M-step uses $N^{(t)}_s$ to estimate $d_s^{(t)}$.
% While the iterative application of the EM guarantees that the likelihood of the estimate converges, it is often the case that the estimates of the parameters converge.  We initialize with the uniform identity coefficients $d_s^{(0)} = \{1/15,1/15,...,1/15\}$.  
As we show in the Supplement, this algorithm is efficient, requiring a small number of iterations before convergence.

%\XXX{Outlines of  the derivation of E and M steps might be helpful too.}

%\subsection{Reducing Spurious Associations}

Inferred relationships can be used to correct association tests for cryptic relatedness thereby reducing spurious, or false-positive, associations.  
This combined algorithm of our estimates of kinship coefficients informing MQLS, we call \emph{pedigree-free MQLS (PFMQLS)}.

\section{Results}\label{sec:results}

There are two categories of results, estimate accuracy and corrections for spurious associations, and simulations that go along with each category.  
We simulated pedigree replicates from the Wright-Fisher (WF) model with parameters: $N$, the number of male and the number of female individuals per generation (meaning there are $2N$ individuals per generation), and $G$, the number of generations.

The data was simulated by holding the pedigree fixed, and drawing an inheritance path for each site from the uniform distribution over inheritance paths.  For each site and inheritance path, the allelic data was drawn using the parameter, $p$---the vector of minor allele frequencies, one for each site.  The founder alleles for site $j$, with possible values in $\{0,1\}$, were each drawn from the Bernoulli distribution with parameter $p_j$.  Once the founder alleles were selected, they were copied along the inheritance path without mutation, so that a descendant inherited a copy of a founding allele from their founding ancestor if there was an undirected path in the inheritance path graph between the descendant and the founder allele.  This simulated data consisted of haplotypes, since it is known which allele was inherited from each parent.

We discuss accuracy in three ways.  Each accuracy measure uses variations on how the genotyped individuals are sampled from the simulation.

\subsection{Improved Estimation of Kinship Coefficients}

To asses the accuracy of the kinship estimates, we simulated pedigree replicates.  For each pedigree, we sampled two extant individuals of interest for which to simulate allelic data.  Recall that the data simulation provides haplotypes.  To obtain the genotypes, the record of each allele's parent was discarded.  The genotype data of the two extant individuals, and not the pedigree, was given to the estimation algorithms.

%We have so far described the simulation method for a single pair of individuals and mentioned which data is given to the estimation algorithms.  
To estimate the allele frequencies, since they vary from site to site, our method requires a set of individuals all simulated using the same allele frequencies.  Therefore, our full simulation produces a number of pairs of individuals, where each pair is simulated from a single pedigree, and all the pedigrees share the same parameter $p$.  Recall that we obtain $p$ using the Laplace estimator applied to independent individuals, one individual from each pedigree.

To assess estimation accuracy, we use a gold standard estimates that are computed during pedigree and inheritance path simulation.  Once the pedigree has been drawn, we apply the algorithm for computing the exact kinship coefficients and the inbreeding coefficients,~\cite{Kirkpatrick2016a}.  Once the inheritance paths have been drawn, we can compute the empirical identity state distribution represented in the data\footnote{Unlike the kinship coefficients, the exact algorithm for obtaining the identity state distribution is exponential, so we use a Monte Carlo approximation similar to~\cite{Sun2014ugrad}.}.  These quantities 1) the kinship and inbreeding coefficients and 2) the identity state distribution, are used as the gold-standard for estimates. 
Table~\ref{table:kinship} shows the results of two estimation methods: the EM algorithm introduced in the Methods section and the covariance-based kinship coefficient estimator introduced by~\cite{Astle2009}.
We found that REAP, which infers kinship from admixed individuals,~\cite{Thornton2012}, does not apply to our setting even when the parameters are set for no admixture, as it found unreasonable kinship and inbreeding coefficients.
%with the assumption that the pair of individuals have the same population allele frequencies, since REAP is a method for inferring kinship from admixed individuals~\cite{Thornton2012}.  However, we conclude that REAP does not apply to our setting, as it provided unreasonable kinship and inbreeding coefficients (many of which were not even probabilities).

%\documentclass[12pt]{article}

%\title{Estimating Identity By Descent}

%\begin{document}
%\maketitle

%% \begin{table}
%% \begin{tabular}{l|l|l|l}
%%   {\bf N} & {\bf G} & {\bf EM Kinship Coeff.} & {\bf MSE IS Distribution}  \\
%%   \hline
%%    3 & 20 & 0.00439916901021821 (3.461485E-03) &  0.0102450176848533 (6.235326E-03) \\
%%   20 &  3 & 0.0058008379781027 (7.545810E-03) &  0.0866195660920303 (7.431484E-02) \\
%%   10 & 10 & 0.00520333004774867 (5.155342E-03) &  0.041054852590465 (2.184016E-02) \\
%%   \hline
%%   {\bf N} & {\bf G} & {\bf Cov. Kinship w/ $p$} & {\bf Cov. Kinship w/ $\hat{p}$} \\
%%   \hline
%%    3 & 20 & 0.0255747390002298 (2.056057E-02) &  11.2363554174777 (3.158361E+01) \\
%%   20 &  3 & 0.0273582050142307 (4.012870E-02) &  13.2571707323805 (3.466623E+01) \\ 
%%   10 & 10 &  0.0214245774102837 (1.630732E-02) &  27.7358534090819 (5.340291E+01) \\
%%   \hline
%% \end{tabular}

\begin{table}
\begin{center}
\begin{tabular}{llllll}
\toprule
\multicolumn{2}{c}{Pedigree}&Id. State&\multicolumn{3}{c}{Kinship} \\
\cmidrule(r){1-2} \cmidrule(r){3-3} \cmidrule(l){4-6}
&&&&\multicolumn{2}{c}{Covariance estimator} \\
\cmidrule(l){5-6}
N&G&EM&EM&with $p$&with $\hat{p}$ \\
\midrule
%3 & 20&  0.024 (1.6E-02)& 0.012 (1.3E-02)& 0.307 (3.8E-02) &  0.340 (2.3E-02) \\
%20 &  3&  0.49 (2.2E-01)& 0.043 (1.4E-02)& 0.0077 (6.8E-03) &  0.77 (2.7E-01) \\
%10 & 10&  0.064 (3.5E-02)& 0.026 (1.3E-02)& 0.025 (1.1E-02) &  0.70 (4.1E-02) \\
3 & 20 & 0.023 (1.42E-02) &  0.012 (1.06E-02) &  0.31 (1.78E-02) &  0.35 (1.67E-02)  \\
10 & 10 & 0.055 (2.53E-02) &  0.016 (1.15E-02) &  0.024 (9.81E-03) &  0.64 (8.15E-02) \\
20 & 3 & 0.45 (1.41E-01) &  0.047 (1.80E-02) &  0.0070 (3.87E-03) &  0.80 (1.92E-01) \\
\bottomrule
\end{tabular}
\end{center}
% \begin{tabular}{l|l|l|l}
%   {\bf N} & {\bf G} & {\bf EM Kinship Coeff.} & {\bf MSE IS Distribution}  \\
%   \hline
%    3 & 20 & 0.0118122857047109 (1.280999E-02) &  0.0238668145035319 (1.619890E-02) \\
%   20 &  3 & 0.0428713150710225 (1.397208E-02) &  0.48921034180392 (2.163708E-01) \\
%   10 & 10 & 0.0262006249341074 (1.302868E-02) &  0.0640110619794971 (3.467485E-02) \\
%   \hline
%   {\bf N} & {\bf G} & {\bf Cov. Kinship w/ $p$} & {\bf Cov. Kinship w/ $\hat{p}$} \\
%   \hline
%    3 & 20 & 0.306723229342686 (3.800315E-02) &  0.339539211196793 (2.303239E-02) \\
%   20 &  3 & 0.00773397760921311 (6.759579E-03) &  0.769917150178498 (2.735859E-01) \\
%   10 & 10 & 0.025399767135831 (1.080048E-02) &  0.699343347328613 (4.148669E-02) \\
%   \hline
% \end{tabular}
\caption{
Given the parameters for the Wright-Fisher (WF) simulation ($N$--the number of male and female individuals per generation, $G$--the number of generations, $p$--the vector of minor allele frequencies), we simulate ten WF replicates each consisting of a pedigree on two extant individuals with data.  For each of these WF pedigree replicates, we simulate genotype data for one hundred independent sites.  The actual kinship coefficients for the two extant individuals and the actual distribution on identity states for the one hundred sampled sites are computed and serve as the gold standard of comparison.  Given just the data, we ran two estimator algorithms to obtain four estimates of either the kinship coefficients or the identity state distribution.    The accuracies of these four estimates relative to the held-out values are shown in the table for several different values of the WF parameters.  The two estimator algorithms are the EM algorithm introduced in this paper (which can estimate both identity states and kinship coefficients) and the covariance-based estimator introduced by Astle and Balding~\cite{Astle2009} (which only estimates the kinship coefficients).  The EM algorithm estimates the kinship coefficients as an expectation over the identity state distribution.  The covariance-based kinship estimator produces two estimates, one based on the actual vector of allele frequencies $p$ and the other for the vector of estimated allele frequencies $\hat{p}$ obtained by empirical estimate as described in Section~\ref{sec:methods}.  While Astle and Balding~\cite{Astle2009} suggested an estimator for $\hat{p}$, in our experience there estimator was unstable and frequently did not converge.
The values in the table show the mean squared error (MSE) between the gold-standard value and the estimated value with the standard deviation in parentheses. Even though the covariance estimator based on the true value of $p$ uses more information than our method, our method still outperforms it in one of the pedigrees and achieves the same performance in another one. When the same amount of information is given to the two methods, our method systematically out-performs the covariance-based method by a large margin.
}
\label{table:kinship}
\end{table}

%\end{document}

\subsection{Comparison with PLINK}
Essentially the worst case of population structure would be if a sample of individuals was from a family, yet they were thought to be unrelated individuals.  Suppose also that the founders of the family are potentially inbred.  Even if the pedigree were known, the pedigree relationships would under-represent the inbreeding since some of the inbreeding occurred chronologically before the known relationships.

\begin{figure}[ht!]
  \begin{center}
  \includegraphics[width=3in]{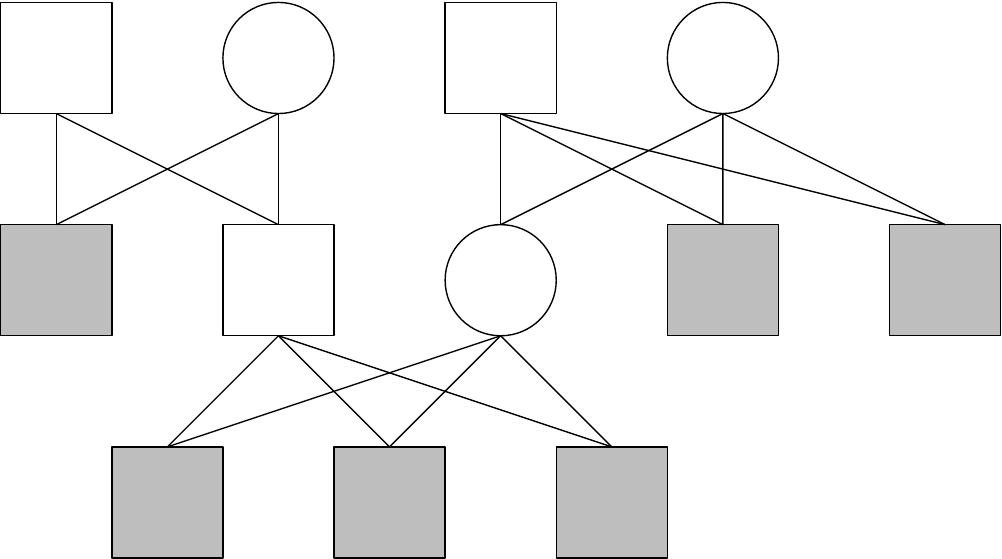}
  \end{center}
  \caption{
{\bf Outbred Pedigree.} This outbred pedigree was used to simulate genotypes from inbred founder haplotypes.  The shaded individuals had genotypes that were typed and used to estimate kinship coefficients with PLINK and the EM method in this paper.
}
  \label{fig:pedigree}
\end{figure}

In order to compare our method to PLINK,~\cite{Purcell2007}, which estimates the kinship coefficients, we simulate just such a scenario.  We use the pedigree Wright-Fisher simulation to produce the founder haplotypes from an inbred population with $N=8$ individuals per generation and $2,4,6,8,...,40$ generations all with $m=500$ sites.  From the most recent generation of the Wright-Fisher pedigree, we draw $4$ founders for an outbred $12$-individual pedigree, see Figure~\ref{fig:pedigree}.  We simulated the recent pedigree genotypes with recombination and considered $6$ of the individuals to have observed genotypes from which we estimated the kinship coefficients.

\begin{figure}[ht!]
  \begin{center}
  \includegraphics[width=3in]{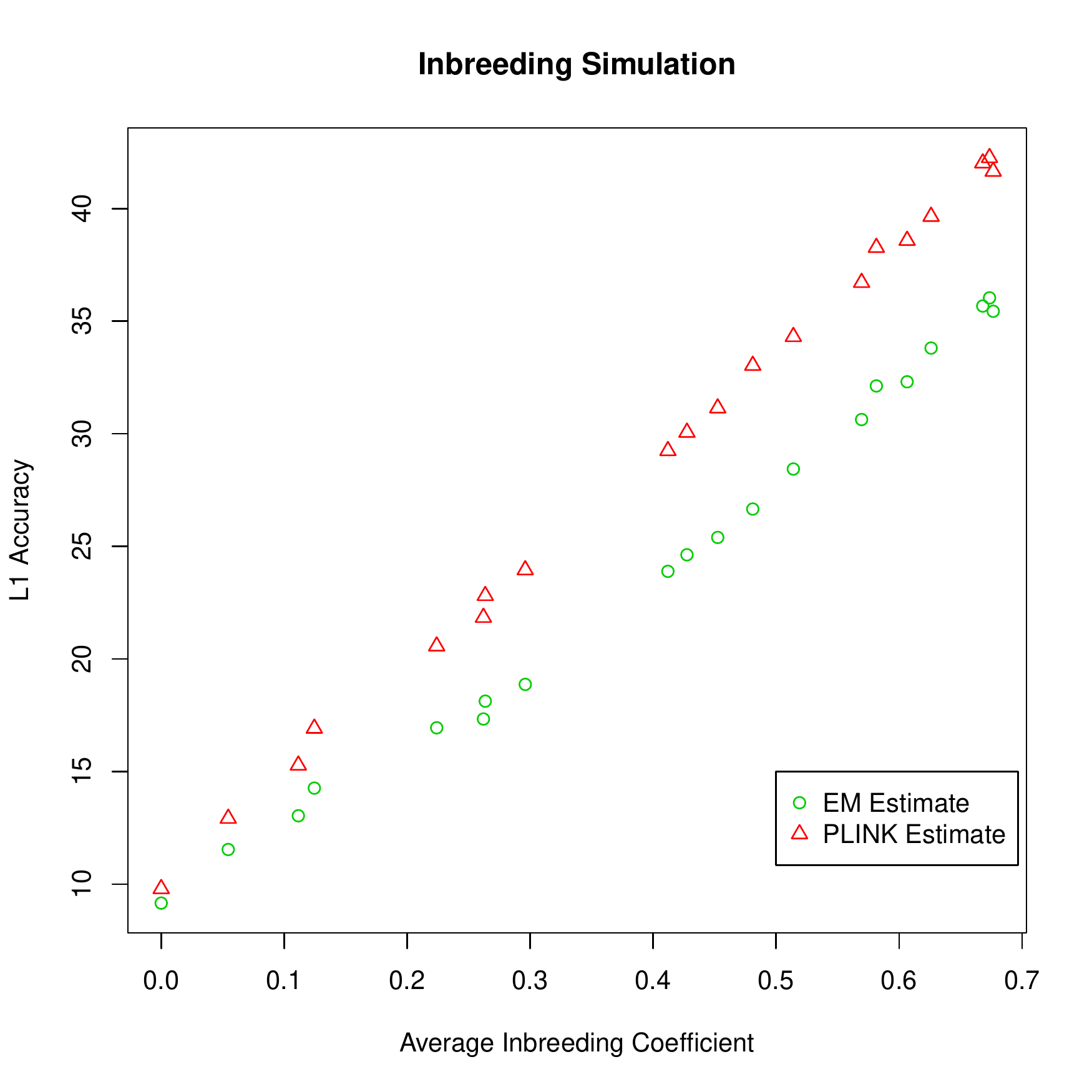}
  \end{center}
  \caption{
{\bf Accuracy of Kinship Estimates.} Comparing kinship estimates of PLINK and our EM method using the $L_1$ accuracy demonstrates that our method has superior accuracy to that of PLINK.  Our method's accuracy margin improves as the amount of inbreeding in the founders increases.
}
  \label{fig:plink}
\end{figure}

To compute the accuracy of the kinship estimates, we found the actual kinship coefficients of the $12$ person pedigree.  This required using a new method of computing kinship coefficients from known founder kinship coefficients, due to~\cite{Kirkpatrick2016a}.  Both methods, ours and PLINK provide estimates of the kinship coefficients.  We computed the sum of the absolute value of the differences between the matrix entries of the estimates and the actual kinship coefficients.  This sum is the $L_1$ accuracy.  In all cases, our method has accuracy far superior to that of PLINK, see Figure~\ref{fig:plink}.

Mathematically, we compare the outbred inference method used by PLINK,~\cite{Purcell2007}, to our identity state approach which considers both outbred and inbred identity states.  Notice that PLINK's approach is limited to considering the 7 outbred identity states which have a ``Yes'' in the first column of Figure~\ref{fig:identstates}.  Our approach considers both inbred and outbred identity states---all states shown in Figure~\ref{fig:identstates}---in a structured learning setting where the identity state for each site is selected along with the frequency for that state.

\begin{figure}[ht!]
  \begin{center}
  \includegraphics[width=3in]{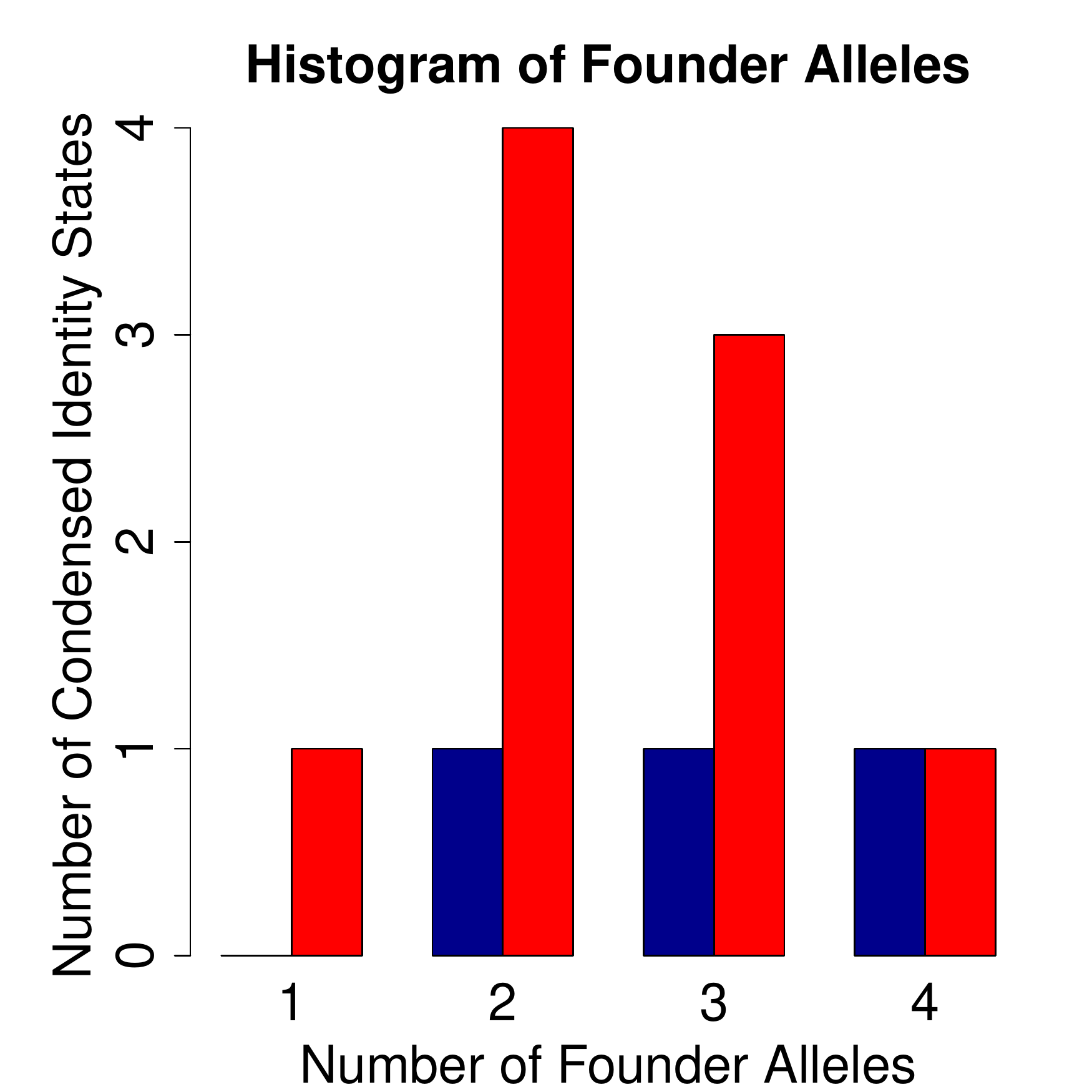}
  \end{center}
  \caption{{\bf Over-fitting.}   The number of founder alleles are on the x-axis, and the y-axis shows the count of the number of condensed identity states with the given number of founder alleles.  The left, blue bars count only the outbred condensed identity states, while the right, red bars count all the condensed identity states (outbred and inbred).  Thus, the red bars are at least the height of the blue bars.  
From this bar plot we can compute the average number of parameters for the outbreeding condensed identity states ($9/3 = 3.00$) and for both inbreeding and outbreeding condensed identity states ($13/9 \simeq 1.44$).}
  \label{fig:overfitting}
\end{figure}

From the bar plot in Figure~\ref{fig:overfitting} we can compute the average number of parameters for the outbreeding condensed identity states ($9/3 = 3.00$) and for both inbreeding and outbreeding condensed identity states ($13/9 \simeq 1.44$).  This shows that on average, PLINK's outbred model will over-fit as compared with our method which selects the best model from both inbred and outbred identity states.

%\begin{figure}[ht!]
%  \begin{center}
%  \includegraphics[width=3in]{100.eps}
%  \end{center}
%  \caption{{\bf Pedigree.} The males are squares and the females are circles.  An individual is shaded gray if they are affected.  Individuals with genotype data are marked with '+'.}
%  \label{fig:pedigree}
%\end{figure}

This over-fitting by PLINK is what we see, when PLINK's $\hat{\pi}$ estimate of kinship nearly recapitulates the kinship computed from an outbred pedigree.  This happens because PLINK explains away the excess homozygosity in the data by outbreeding which introduces many independent founder alleles---the founder alleles all appear in distinct connected components of the identity states.
On the other hand our approach discovers that outbreeding explains the excess homozygosity less well than inbreeding, because the outbreeding explanation has a lower likelihood than an explanation involving inbreeding.  Therefore our approach estimates, from the data, kinship coefficients that deviate from that predicted by the (outbred) pedigree structure, precisely because inbreeding among the founders provides a better model.  Indeed, under the setting with a lot of inbreeding among the founders of a pedigree, an outbred-only model like that used by PLINK might have a significantly lower likelihood than a model that allows inbreeding as an explanation for the observed excess homozygosity.

\subsection{Fewer Spurious Associations}

Similar to the simulations in the previous section, we simulated pedigree replicates from the Wright-Fisher (WF) model.
Unlike the previous simulations, we sampled $k$ extant individuals and discard the pedigree graph.  The genotypes at site $j$ were simulated as before by drawing the founder alleles uniformly at random from the population distribution which is Bernoulli with parameter $p_j$, and then inheriting those alleles along the edges indicated by the inheritance path.The case-control simulation then involves two pedigree replicates which share the same founder allele frequencies.   The presence of the two sets of family relationships confounds most association tests and results in very low power. 

Each site was independently taken to be the disease site, resulting in $m$ true positive tests per simulated pair of pedigrees.  The affection status of each individual was computed from the genotype assuming an almost recessive trait and the minor allele to be the disease allele.  Assuming the minor allele is $0$, the penetrance probabilities for the disease given the genotype of person $i$ were $\bbP(D|G^i=0)=0.95$ and $\bbP(D|G^i=1)=\bbP(D|G^i=2)=0.05$.  Several association tests were applied to the simulated genotypes of the two pedigrees: 
%the $\chi^2$ test, 
the Cochran-Armitage trend test, 
the ROADTRIPS RM test,~\cite{Thornton2010}, and the PFMQLS test which is the MQLS test given the kinship coefficients estimated by the EM algorithm. 
%\footnote{We also tried comparisons with the method REAP of \cite{Thornton2012}, however we were unable to obtain reasonable inferences on kinship coefficients in our setup with this method, presumably because REAP is designed with a different setting in mind, namely admixed individuals.}

The simulations where conducted with the following parameters: $N=50$ number of male/females per generation, $G=25$ number of generations, $k=10$ individuals sampled, and $m=400$ sites.  We suggest using the Bonferroni threshold.\footnote{Our association results are demonstrated with p-values, but our method is equally amenable to the use of false discovery rates and the computation of q-values.}  Overall, we find that the Bonferroni threshold can favor the PFMQLS test, see Supplement.

\begin{figure}[ht!]
  \begin{center}
  \includegraphics[width=3in]{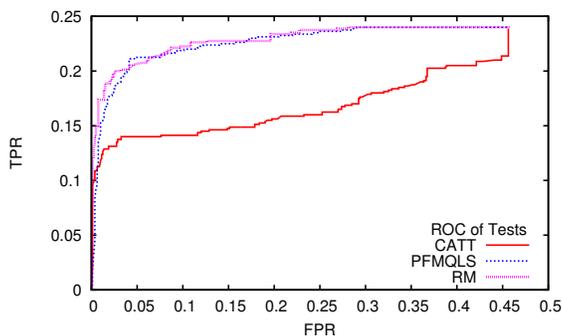}
  \end{center}
  \caption{{\bf ROC plot.} The x-axis is the false positive rate and the y-axis is the true positive rate.  There are several tests shown:
%, the $\chi^2$ test (CHISQ), 
the Cochran-Armitage trend test (CATT),
the pedigree-free MQLS test (PFMQLS), 
and the ROADTRIPS RM test.  The performances of RM and PFMQLS are almost indistinguishable and far superior to that of CATT.
}
  \label{fig:roc}
\end{figure}

 We summarized the data in a receiver operating characteristic (ROC) plot.  The $(x,y)$ points for the ROC plot are the false positive rate (FPR) and true positive rate (TPR) for a particular p-value threshold for the test.  By considering multiple thresholds, we can look across all the simulations to find the FPR for all the non-disease site and to find the TPR for all the disease sites, see Figure~\ref{fig:roc}.  
%There is one disease site per simulation, so it is difficult to determine whether a TPR is significant.
While there is a slight difference between the performance of the PFMQLS and ROADTRIPS, the number of simulations suggests that this difference may not be significant.  The performance of the Cochran-Armitage trend test (CATT) is very poor due to many false positives.  Both PFMQLS and ROADTRIPS avoid the spurious false positives.   We conclude that PFMQLS is as good as the state-of-the-art represented by ROADTRIPS while providing an intuitive and interpretable model of relatedness.

\section{Conclusion}
We present a method for inferring the kinship coefficients that relies on identity states---a more detailed description of the pedigree than used by most kinship inference methods.  Our method is an EM algorithm that infers the identity state distribution without assuming a known pedigree.  The accuracy of our method depends on the number of sites and is reliable with as few as 64 independent sites as input\footnote{From the whole genome of correlated sites, it is feasible to extract many more than 64 independent sites for input to our method.}.  Our results show that our kinship estimates out-perform the covariance kinship method and other recent methods for kinship estimation by a large margin.  Our EM kinship estimates can also correct an association test to produce state-of-the-art accuracy.  
%Our method treats sites independently, so an open problem is to design an efficient method for inferring identity states which takes advantage of correlated sites. 
%correlated sites can always extract independent sites as input to our method, and since there are thousands of independent sites in the genome, obtaining at least 64 for input to our method is quite feasible.

Constructing the kinship matrix with our method requires a pair-wise comparison of individuals' genomes.  Such an approach can be computationally intensive, and future work includes a method avoiding this quadratic cost.  A potential drawback of using an EM algorithm is that it finds local optima: our current results each use a single run of EM, and so random restarts could potentially improve the results.

Other areas of future work involve simulating complex diseases which produce the disease trait by interaction of multiple sites in the genome.  The results presented here were for a simple nearly recessive disease which probably accounts for the high accuracy of PFMQLS and ROADTRIPS as seen in the area under the curve for the ROC plot.  In addition to simulating complex diseases, future work involves tailoring a test to the setting of using kinship corrections to detect epistasis.

%\XXX{we should acknowledge the limitations of our method. For example, we consider only pairwise interaction between individuals, considering all pairs can be computationally intensive, etc.  FDR is orthogonal and interchangable with p-values. EM local optima... better with random restart.}

%\XXX{future work, use recombination, more complex disease, custom test?}

%\section*{Acknowledgments} 

\section*{Author Disclosure Statement.}
BK is the owner of Intrepid Net Computing.

\bibliographystyle{humanbio}
\bibliography{pedbib}

\newpage
\section*{Supplement: Correcting for Cryptic Relatedness in Genome-Wide Association Studies}

\subsection{Reducing Spurious Associations}

Inferred relationships can be used to correct association tests for cryptic relatedness thereby reducing spurious, or false-positive, associations.  Notably, the MQLS,~\cite{Thornton2007} test relies on kinship coefficients calculated from a known pedigree to correct for the dependencies caused by relatedness that would confound tests that assume independence, such as the $\chi^2$ test.  We propose to reduce spurious associations in data sets having an unknown pedigree by using our EM algorithm for estimating the kinship coefficients for every pair of individuals.  
Recall that for every pair of individuals, we obtain estimates of the inbreeding coefficients of each individual and the kinship coefficient between them.  
For each of the $N$ individuals, we will have $N-1$ estimates of the inbreeding coefficient which we average to obtain a single estimate.  This leaves us with a matrix of estimates with the off-diagonals being estimates of the kinship coefficients and the diagonal being the estimates of the inbreeding coefficients.  
%Since we wish to provide this matrix to MQLS, we also add 1 to all the diagonal entries in the matrix and take the kinship matrix as the closest positive semi-definate matrix~\cite{Higham1988}.  
%However, this matrix may not be positive semi-definite (since we only consider pair-wise interactions between individuals) which is a requirement of input to MQLS.  Therefore, we find the closest positive semi-definite matrix,~\cite{Higham1988}---a valid MQLS kinship matrix.

This combined algorithm of our estimates of kinship coefficients informing MQLS, we call \emph{pedigree-free MQLS (PFMQLS)}.  While it is possible to tailor-design a test based on these EM kinship coefficients, this approach of running MQLS on our EM results allows us to judge whether the kinship coefficients estimated by the EM algorithm can successfully reduce spurious associations.

\subsection{Results: Improved Estimation of Kinship Coefficients}

\begin{figure}[ht!]
  \begin{center}
  \includegraphics[width=3in]{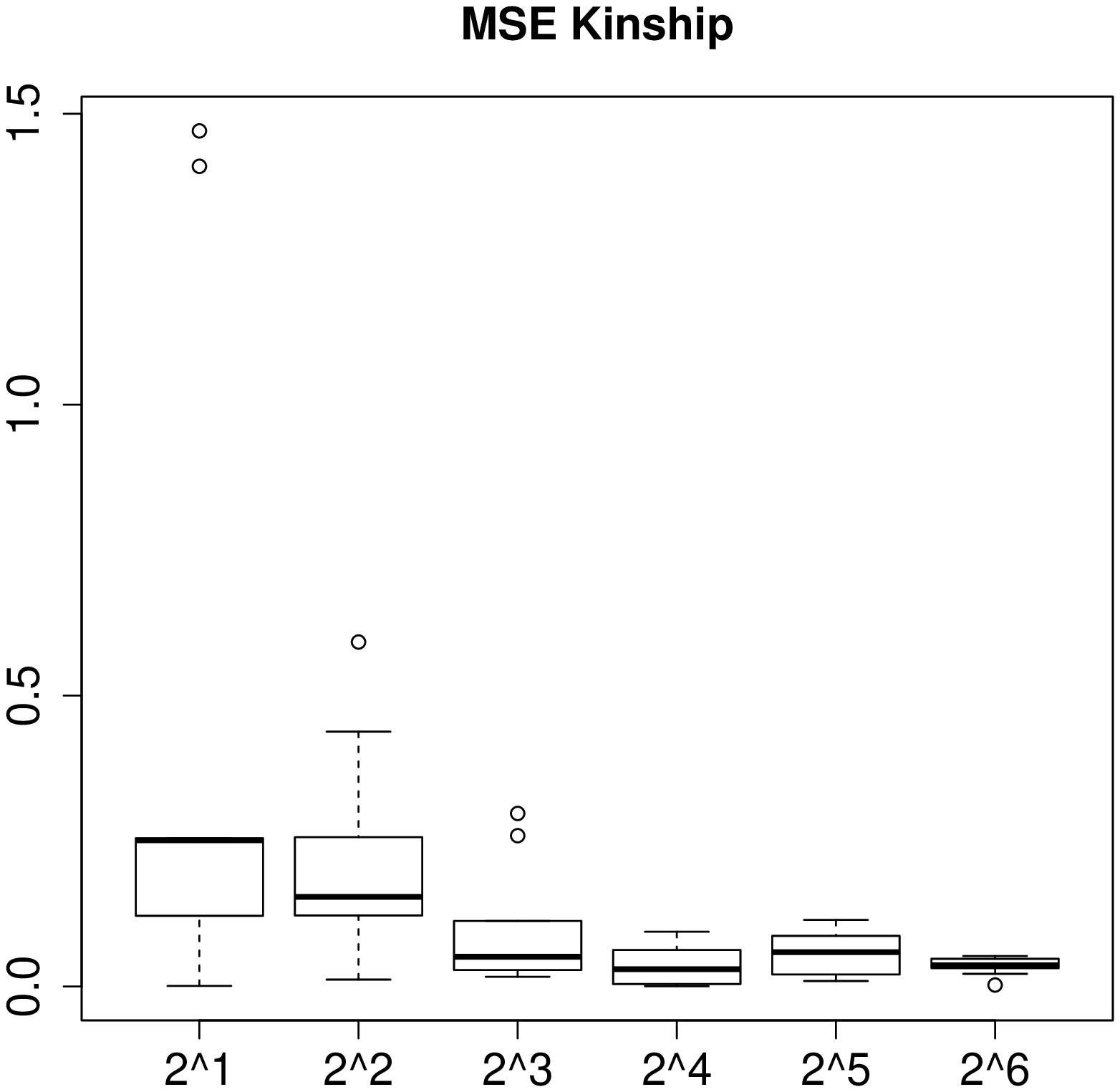}
  \includegraphics[width=3in]{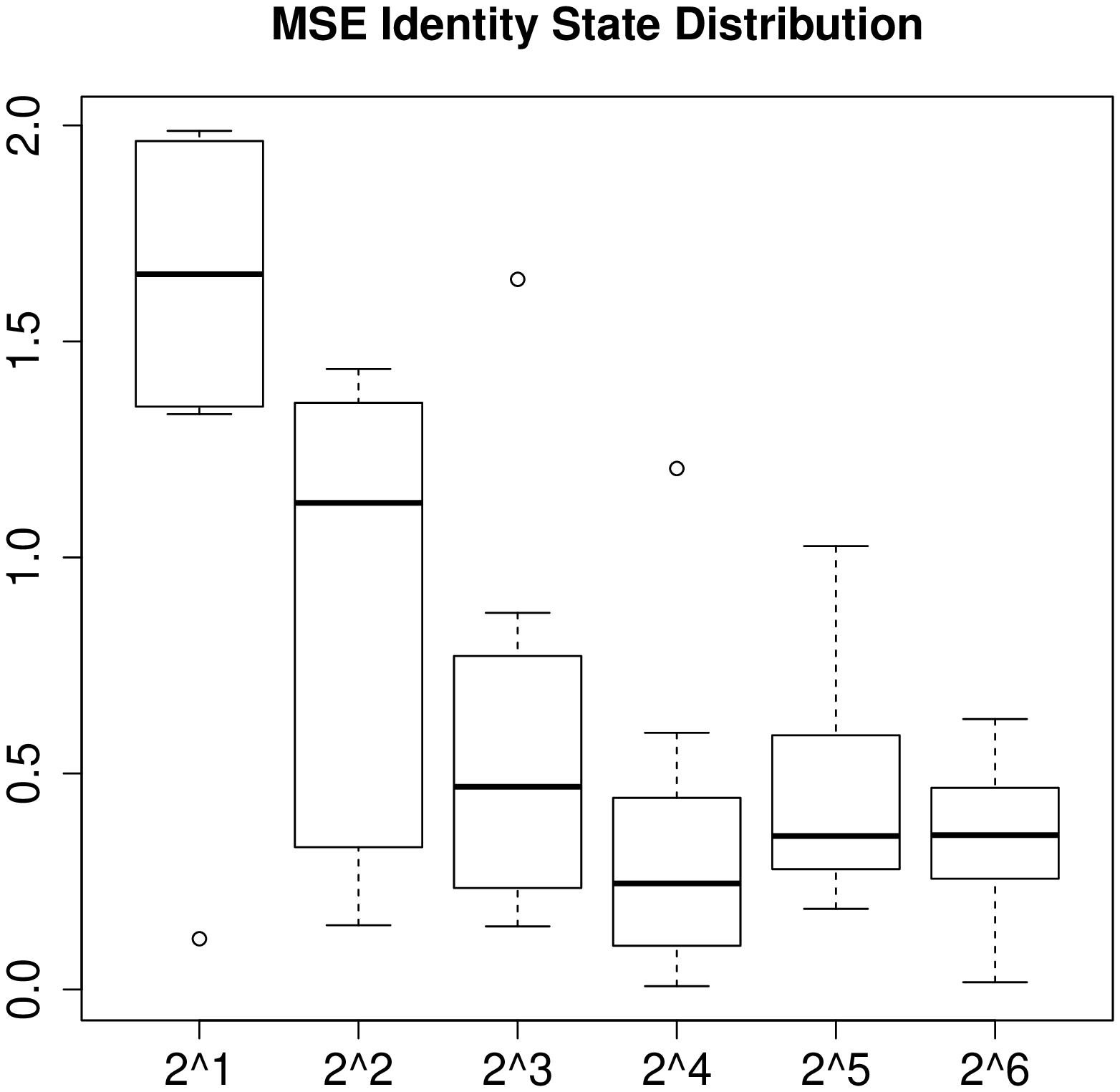}
  \end{center}
  \caption{{\bf The number of sites and the accuracy of the estimates.} As the number of sites increase, log-scale on the x-axis, the accuracy, on the y-axis of both the kinship and the identity state estimates, improves.  The maximum number of sites shown here is $2^6 = 64$.}
  \label{fig:sites}
\end{figure}

\begin{figure}[ht!]
  \begin{center}
  \includegraphics[width=3in]{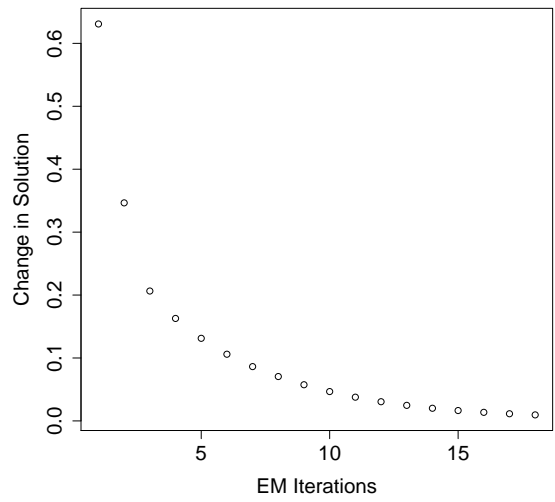}
  \end{center}
  \caption{{\bf $L_1$ distances between the EM solutions of successive iterations.}
For a single pair of individuals, the figure records the $L_1$ distance between every pair of solutions for successive iterations.  As the iterations, x-axis, increase the $L_1$ distances, y-axis, decrease rapidly towards zero.}
  \label{fig:l1}
\end{figure}

Figure~\ref{fig:sites} shows the effect of the number of sites on the accuracy of the estimates, both in terms of the kinship coefficient estimates and of the identity state distribution estimate.  %The values on the x-axis are powers of two which indicate the number of sites used for that set of 10 simulations.  
The simulations were performed exactly as they were for the table~(see paper) with the actual allele frequencies generated uniformly and the estimated allele frequencies being obtained empirically from independent individuals in the simulation.  The estimated allele frequencies were then used to estimate the identity state distribution and the kinship coefficients.
%Considering the efficiency of our implementation of the EM algorithm, w
We show that it is possible to use only a few EM iterations to obtain a stable solution, Figure~\ref{fig:l1}.  
%It is readily seen that the EM solution quickly converges as the number iterations increases, since the $L_1$ distance drops towards zero. 

\subsection{Results: Fewer Spurious Associations}

\begin{figure}[ht!]
  \begin{center}
  \includegraphics[width=5in]{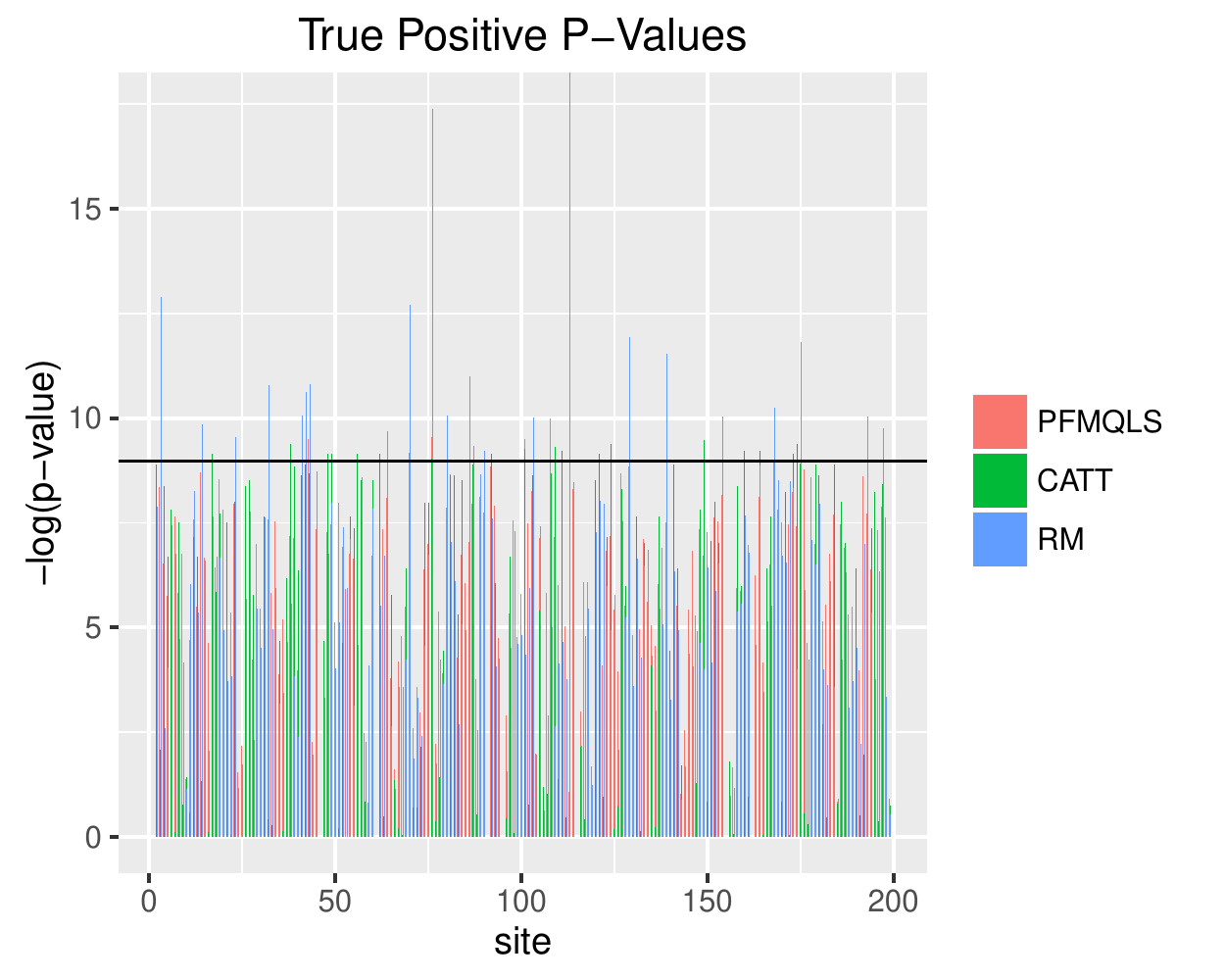}
  \end{center}
  \caption{{\bf True positives.} On the x-axis are the sites, and the y-axis the negative log of the p-value.  We show here the negative log p-value for the pedigree-free MQLS test (PFMQLS),
%, the $\chi^2$ test (CHISQ), 
the Cochran-Armitage trend test (CATT),
and the ROADTRIPS RM test.  
The horizontal line in the figure gives the Bonferroni-corrected threshold for a site-specific significance of $0.05$.  Any spikes that protrude above the line indicate sites that are significant and true positive.}
  \label{fig:tppval}
\end{figure}

\begin{figure}[ht!]
  \begin{center}
  \includegraphics[width=5in]{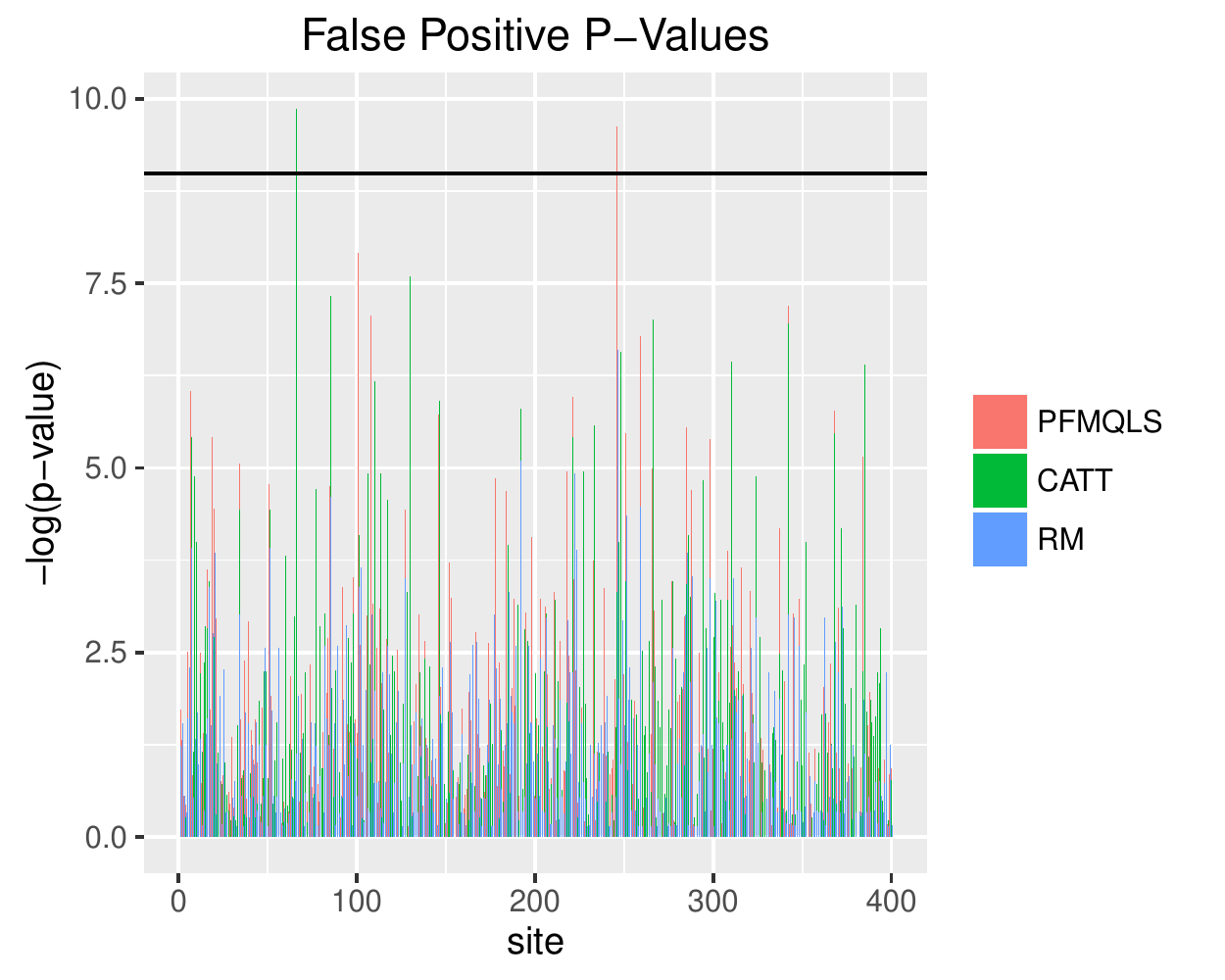}
  \end{center}
  \caption{{\bf False positives.} On the x-axis are the sites, and the y-axis the negative log of the p-value.  We show here the negative log p-value for the pedigree-free MQLS test (PFMQLS), 
%, the $\chi^2$ test (CHISQ), 
the Cochran-Armitage trend test (CATT), 
and the ROADTRIPS RM test.  
The horizontal line in the figure gives the Bonferroni-corrected threshold for a site-specific significance of $0.05$.  Any spikes that protrude above the line indicate sites that are significant and false positive.
}
  \label{fig:fppval}
\end{figure}

We ran a simulation as described in the paper with $m=400$ sites.  Figure~\ref{fig:tppval} shows the true positive p-values while Figure~\ref{fig:fppval} shows the false positive p-values.\footnote{Our association results are demonstrated with p-values, but our method is equally amenable to the use of false discovery rates and the computation of q-values.}  Overall, these figures illustrate that the Bonferroni threshold can favor the PFMQLS test.

\end{document}